\documentclass[lettersize,journal]{IEEEtran}
\usepackage{amsmath,amsfonts}
\usepackage{algorithmic}
\usepackage{algorithm}
\usepackage{array}
\usepackage[caption=false,font=normalsize,labelfont=sf,textfont=sf]{subfig}
\usepackage[justification=centering]{caption}
\usepackage{textcomp}
\usepackage{stfloats}
\usepackage{url}
\usepackage{verbatim}
\usepackage{graphicx}
\usepackage{cite}
\usepackage{multirow}
\usepackage{chngcntr}%图片重新编号需要调用这个宏包
\counterwithout{figure}{section}%取消图片按章节编号
\begin{document}

\title{	Human Activity Recognition with Low-Resolution Infrared Array Sensor Using Semi-supervised Cross-domain Neural Networks \\for Indoor Environment}

%\author{San Zhang, Si Li, Wu Wang

\author{Cunyi Yin, Xiren Miao, Jing Chen, Hao Jiang, Deying Chen, Yixuan Tong and Shaocong Zheng
% IEEE Publication Technology,~\IEEEmembership{Staff,~IEEE,}
        % <-this % stops a space
\thanks{This work was supported by the Natural Science Foundation of Fujian Province, China, grant number 2022J01566 and the China Scholarship Council, China, grant number 202206650012.     {\itshape (Corresponding author: Jing Chen)}

The authors are with the College of Electrical Engineering and Automation, Fuzhou University, Fuzhou 350108, China (e-mail: 200110004@fzu.edu.cn (C.Y.); mxr@fzu.edu.cn (X.M.); chenj@fzu.edu.cn (J.C.); jiangh@fzu.edu.cn (H.J.); 200120067@fzu.edu.cn (D.C.); 200127027@fzu.edu.cn (Y.T.); 210127054@fzu.edu.cn (S.Z.)).
}}% <-this % stops a space
%\thanks{Manuscript received November 1, 2021; revised November 20, 2021.}}

%\markboth{Journal of \LaTeX\ Class Files,~Vol.~14, No.~8, September~2022}%
\markboth{}%
{Shell \MakeLowercase{\textit{et al.}}: Human Activity Recognition with Low-Resolution Infrared Array Sensor Using Semi Supervised Cross-Domain Neural Networks for Indoor Environment}


\maketitle

\begin{abstract}
Low-resolution infrared-based human activity recognition (HAR) attracted enormous interests due to its low-cost and private. 
In this paper, a novel semi-supervised crossdomain neural network (SCDNN) based on 8 $\times$ 8 low-resolution infrared sensor is proposed for accurately identifying human activity despite changes in the environment at a low-cost. 
The SCDNN consists of feature extractor, domain discriminator and label classifier. 
In the feature extractor, the unlabeled and minimal labeled target domain data are trained for domain adaptation to achieve a mapping of the source domain and target domain data. 
The domain discriminator employs the unsupervised learning to migrate data from the source domain to the target domain. 
The label classifier obtained from training the source domain data improves the recognition of target domain activities due to the semi-supervised learning utilized in training the target domain data. 
Experimental results show that the proposed method achieves 92.12\% accuracy for recognition of activities in the target domain by migrating the source and target domains. 
The proposed approach adapts superior to cross-domain scenarios compared to the existing deep learning methods, and it provides a low-cost yet highly adaptable solution for cross-domain scenarios. 
\end{abstract}

\begin{IEEEkeywords}
human activity recognition (HAR), low-resolution infrared array sensor, cross-domain recognition.
%Article submission, IEEE, IEEEtran, journal, \LaTeX, paper, template, typesetting.
\end{IEEEkeywords}

\section{Introduction}

%\IEEEPARstart{T}{his} file is intended to serve as a ``sample article file''
%for IEEE journal papers produced under \LaTeX\ using
%IEEEtran.cls version 1.8b and later. The most common elements are covered in the simplified and updated instructions in ``New\_IEEEtran\_how-to.pdf''. For less common elements you can refer back to the original ``IEEEtran\_HOWTO.pdf''. It is assumed that the reader has a basic working knowledge of \LaTeX. Those who are new to \LaTeX \ are encouraged to read Tobias Oetiker's ``The Not So Short Introduction to \LaTeX ,'' available at: \url{http://tug.ctan.org/info/lshort/english/lshort.pdf} which provides an overview of working with \LaTeX.
\IEEEPARstart{T}{he} era of the Internet of Things (IoT) arrived, the elements of smart and convenient can be common sight in people’s life \cite{2018Sensor}. 
Human Activity Recognition (HAR) system, an import part of IoT technology, become one of the research hotspots in recent years. 
The tasks such as power operation and factory operation, some dangerous activities can be detected by HAR system to protect the safety of workers or avoid accidents. 
In a smart home scenario, human utilize HAR system to interact with electrical equipment or monitor their health.

The existing approaches to achieve HAR mainly resort to computer vision and kinds of sensors. The computer visionbased approaches \cite{Deniz2015Recognition,2015A,2015Human,2017Recognizing,2019An,2020A} were common and widespread way of implementing HAR, however, reliance on cameras to acquire data is heavily influenced by light and obscuration. 
More importantly, smart homes were often less receptive to the approach due to privacy concerns. 
Researchers utilized smartphones or wearable devices to recognize human activities in \cite{SAN2017186,2016Deep,2019Recognizing,2016Human,Mohammed2018A,2020Deep}, but they were not always carrying equipment. 
Channel state information (CSI)-based HAR methods \cite{2014E,2015Anti,2018FallDeFi} received a lot of attention in recent years due to its nonintrusive sensing properties. 
The CSI signal acquisition requires more than two devices of intel 5300NICs or Atheros, both of them need a serial connection to a PC for data acquisition. 
It is not a convenient solution in terms of either device acquisition or deployment. 

Recently, infrared sensors are proposed as a new HAR schemes because of its convenient deployment and lightindependent \cite{2020Multiple,2018A,2019Action}. 
Low-resolution infrared sensors, in particular, are cheaper than higher-resolution sensors and offer privacy protection. 
Jeong $et$ $al.$ \cite{2014Probabilistic} interpolated 8 $\times$ 8 low-resolution infrared sensor data and utilized the higher dimensional data obtained from the interpolation to recognize human activity. 
Mashiyama $et$ $al.$ \cite{2015Activity} proposed a method for activity recognition utilizing a AMG8833, a 8 $\times$ 8 low-resolution infrared array sensor. 
The temperature distribution obtained from the sensor is analysed and classified into five basic states, no event, stopped, walking, sitting and emergency.
Burns $et$ $al.$ \cite{2019Fusing} employed two 32 $\times$ 31 infrared sensors for person behaviour detection from both the top and side directions. 
The limited dimensionality of the data collected by low-resolution sensors makes feature extraction more difficult. 
It is doubtless a great challenge to achieve high accuracy for low-resolution infrared array sensors. 
With the addition of deep learning, it is possible to employ low-resolution infrared to achieve higher accuracy of HAR. 
Fan $et$ $al.$ \cite{2017Robust} utilized an 8 $\times$ 8 infrared array sensor with a deep learning approach to detect the activity of human falling. 
Munkhjargal $et$ $al.$ \cite{8616393} proposed a device-free and unobtrusive indoor human pose recognition system based on low-resolution infrared sensors and deep convolutional neural networks (DCNN). 
They also utilized DCNN and low-resolution infrared sensors recognized yoga poses to provide an IoT-based yoga training system for yoga practitioners \cite{8707064}. 
Shih $et$ $al.$ \cite{9111825} fused multiple low-resolution infrared images and employed a deep learning algorithm to recognize human activity. 
Nevertheless, there are some limitations with these methods. 
The traditional deep learning requires massive amounts of data for training, which is labor intensive for data collection. 
The traditional deep learning models were solidified, and the original models no longer were applicable to the new environment when the environment changed. 
The data must be re-collected and the model must be re-trained in the new environment. 
Yin $et$ $al.$ \cite{article} acquired and filtered low-resolution infrared data from different environments to reduce the effect of environmental changes on the data. 
The approach still had to be trained in large amounts of data and cannot essentially find a solution to the impact on environmental changes. 
The cross-domain problem caused by environmental changes became a key constraint on the utilize of low-resolution infrared for HAR. 
Cross-domain research achieved well results in the field of imaging, and transfer learning is hot in cross-domain. 
Finn $et$ $al.$ \cite{Finn2017ModelAgnosticMF} selected data from a new environment to train the parameters of the model explicitly, allowing a small number of gradient steps to produce excellent generalization performance on the task. 
Lima $et$ $al.$ \cite{7829262} extracted deep knowledge of CNN models from large datasets and then transfered it by fine-tuning to a limited number of remotely sensed (RS) images on a marine frontier identification task. 
Alshalali $et$ $al.$ \cite{8947855} utilized a pre-trained CNN model and fine-tuned the model based on an extreme learning machine (ELM) for target recognition and image classification applications.
TL-GDBN, a growing dynamic Bayesian network with transfer learning, was proposed by Wang $et$ $al.$ \cite{8478799} to initialize and pre-train the basic deep belief network (DBN) structure with a single hidden layer to accelerate the learning process and improve model accuracy. 
Tao $et$ $al.$ \cite{Tao2019Active} fine-tuned the pre-trained hierarchical stacked sparse self-coding (SSAE) network and limited training samples from the source domain to the target domain utilizing a limited number of labeled samples selected from the source and target domains via an active learning strategy. 
Mohammadi $et$ $al.$ \cite{9311169} employed infrared cameras for collecting human activity data and proposed a sleep monitoring method incorporating supervised learning mechanisms in transfer learning to analyse human posture and movement. 
The transfer learning approaches described above solved the cross-domain problem to a certain extent, but they were model migration approaches that requiring supervised learning of data collected in a new environment to achieve model migration. 
The method to model migration, with a large amount of data to be collected and annotated in the new environment, is not an efficient and cost effective method. 
Ganin $et$ $al.$ \cite{2017Domain} proposed the domain adversarial neural network (DANN) to study minist datasets in images across domains. The approach to training unlabeled data in new environments to achieve data migration provided a new idea of cross-domain research.

In this paper, a novel semi-supervised cross-domain neural network (SCDNN) is proposed for HAR utilizing a low-resolution infrared array sensor. 
Only one 8 $\times$ 8 low-resolution infrared array sensor is utilized in the signal acquisition. 
The employment of low-resolution infrared sensors both reduces hardware costs and protects human privacy. 
The DANN is applied in low-resolution infrared for HAR to adapt the changing environment. 
With the training of DANN, the infrared signals of the new environment (target domain) and the original environment (source domain) are mapped into the same feature space, and the classifier trained on the source domain is directly employed to classify the infrared signals in the target domain. 
The SCDNN is an improvement on the DANN. 
The difference is the cross-domain data migration method in SCDNN employs unlabeled and a tiny minority labeled target domain data for data migration. 
Unsupervised and semi-supervised learning mechanisms are leveraged in SCDNN to perform domain adaptation training on source and target domain data. 
The proposed method achieves accurate recognition of human activities in the target domain.
We summarize the main contributions of our work as follows:
\begin{itemize}
\item{A cross-domain low-resolution infrared HAR method is proposed. 
In contrast to traditional deep learning methods applied to source domain data exclusively. 
The source and target domain data are mapped into a unified feature space. 
A high accuracy HAR result is obtained for a novel environment. 
The problem of degradation of recognition accuracy due to environmental changes in traditional methods is tackled.}
\item{The proposed HAR approach is a low-cost training method. 
A large amount of data acquisition and annotation work is not essential in the novel environment. 
There is a small-scale unlabeled and tiny minority labeled data is necessary for the adaptive training.}
\item{The idea of cross-domain is applied to HAR in low-resolution infrared. 
A semi-supervised learning mechanism is employed in training. 
Superior spatial feature mapping is achieved for the source and target domain data.
The SCDNN model trained with small-scale target domain data achieves accurate recognition for human activities in the novel environment.}
\item{A single 8 $\times$ 8 low-resolution infrared sensor is utilized to collect data. There are 8 categories of activities classified accurately, which cover static, dynamic activities and similar activities. A recognition accuracy of 92.12\% is achieved in cross-domain scene.}
\end{itemize}

The rest of this paper is organized as follows. 
In Section II, we describe the proposed principle of HAR method based on SCDNN. 
The experimental results and analysis are shown in section III. 
Section IV draw conclusions from the results and discuss about our future work plan.

\section{SYSTEM OVERVIEW}
The system framework of the proposed cross-domain based low-resolution infrared HAR approach is shown in Fig. \ref{fig_1}. 
The proposed method consists of 3 steps, infrared signals acquisition and processing, model training and activity recognition. 
The signals acquisition is divided into source and target domain data acquisition. 
The data collected in scene I is employed as the source domain data and the scene II is utilized as the target domain data. 
The data is collected and sent to the server for data processing. 
The same data segmentation and filtering is performed on the source and target domain data.
The source domain data are all labeled and divide into a training set and a validation set. 
A small-scale target domain data is selected as the training set and the rest as the test set.
Only a tiny minority training set is labeled and the remainder is left unlabeled.
\begin{figure*}[!t]
	\centering
	\includegraphics[width=7.2in]{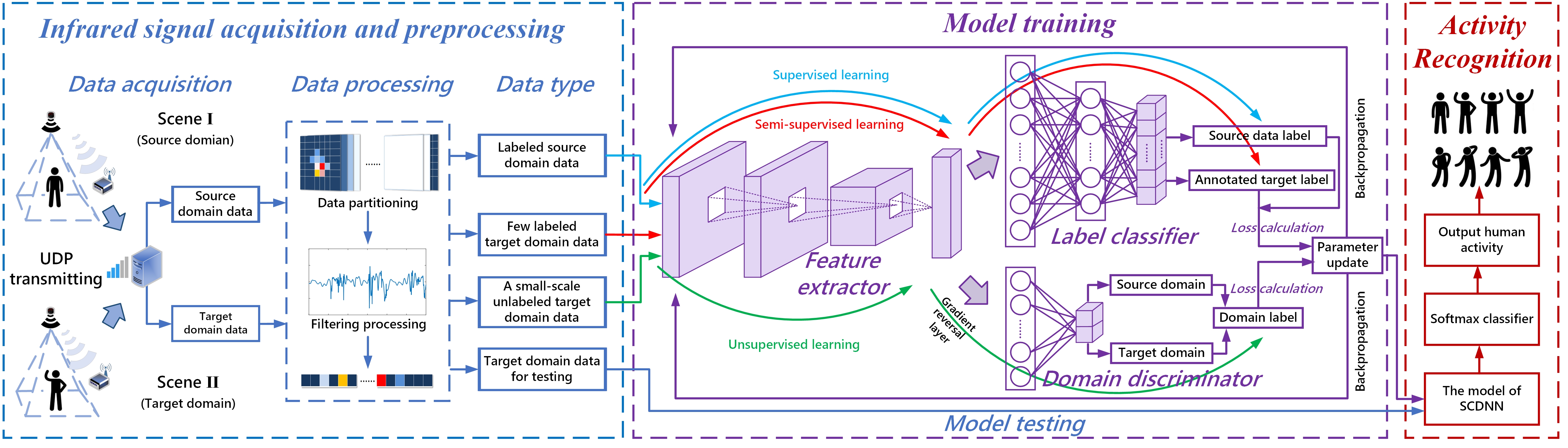}%
	\caption{Schematic diagram of HAR method based on SCDNN.}
	\label{fig_1}
\end{figure*}

The source domain data is trained in SCDNN model for supervised learning. 
The validation set of the source domain optimizes the label classifier in training to obtain superior results in classifying human activities. 
The source domain data and the unlabeled target domain data are fed into the domain discriminator for unsupervised learning training, with the aim of making the target domain data closer to the source. 
The label classifier learns the distribution features of the target domain data as the labeled target domain data is learned supervised. 
The aim of the training is the label classifier can discriminate the category of the source domain data, and the domain discriminator unable to distinguish the data source. 
If it is difficult for the domain discriminator to separate the data sources, the source and target domain data are well mapped into the same space. 
Test data from the target domain is mapped in the feature extractor and fed into the label classifier to classify the activities of human.

\section{METHODOLOGY}
\subsection{Data Collection}
The Grid-EYE AMG8833, an 8 $\times$ 8 low-resolution infrared array sensor manufactured by Panasonic, is employed in the person behaviour recognition method proposed in this paper. 
The appearance and structure of the AMG8833 can be referred to in Fig. \ref{fig_2}. 
It contains a total of 64 thermocouples in 8 horizontal and 8 vertical directions. 
The temperature range for detection is between -20$^{\circ}$C and 80$^{\circ}$C with a tolerance of ±2.5$^{\circ}$C. 
The AMG8833 is capable of detecting objects within a range of 7m.
\begin{figure}[b]
	\centering
	\setcounter{figure}{1}
	\includegraphics[width=2.5in]{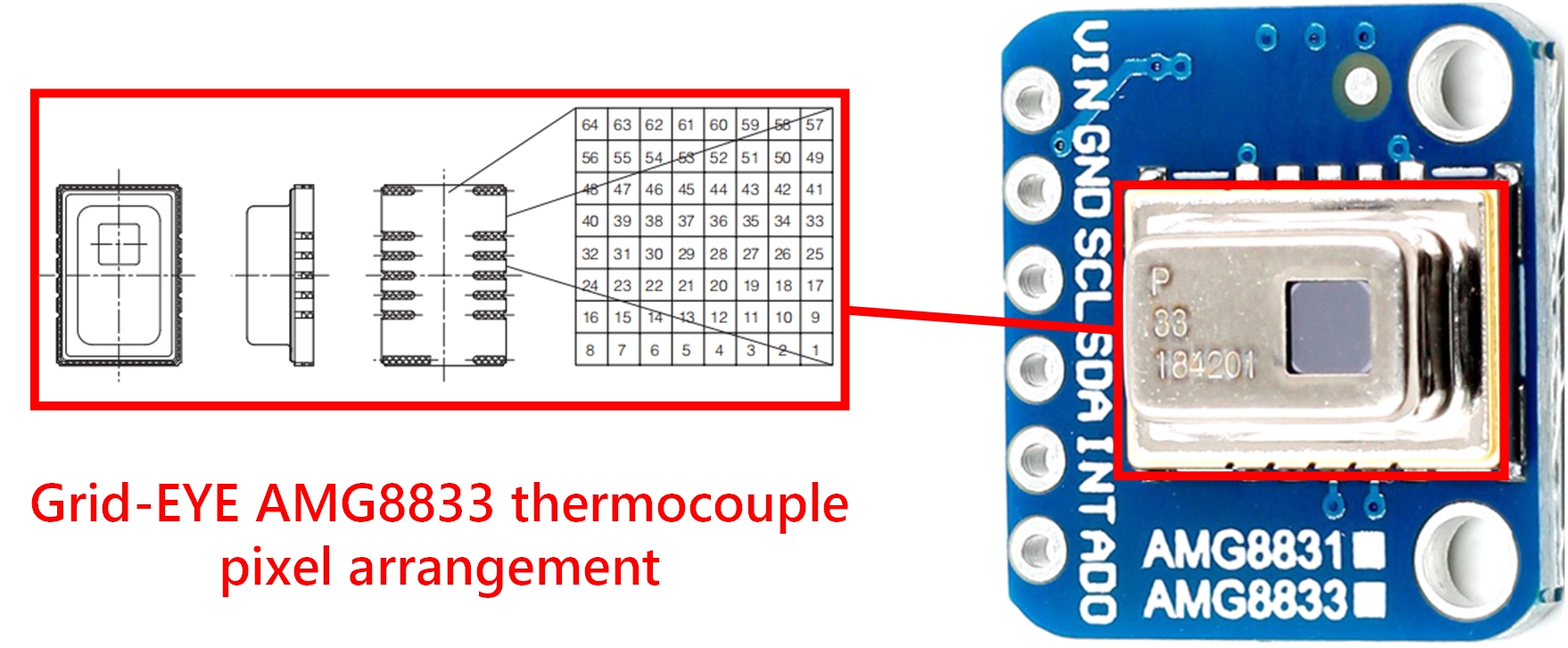}
	\caption{The 8$\times$8 pixel low-resolution infrared array sensor of the Grid-EYE AMG8833 module.}
	\label{fig_2}
\end{figure}
An ESP8266 is utilized as a microprocessor to obtain the measured temperature values via an I$^{2}$C bus connection with the AMG8833. 
The infrared signals of different activities are captured by the AMG8833 and sent to the ESP8266.
Infrared signals are sent to the server via UDP utilizing the WiFi serial communication of the ESP8266.  
The microprocessor sends the infrared signals out at a rate of m frames/s. 
Each frame of the infrared signal is an 8 $\times$ 8 matrix. 
The infrared signals are captured in 2 separate scenarios, scene I and scene II.
The signals captured in the 2 scenes are employed as source domain data and target domain data respectively.

\subsection{Data Preprocessing}
Human activities occur over a period of time. 
The timeseries infrared signal is normally utilized to reflect human activities. 
The dimensions of different activity samples needs to be consistent during the training process. 
In this work, the length of each sample is set to m frames. 
The acquisition process of the infrared signal is continuous, but each behaviour is not necessarily divisible by m after L consecutive frames are acquired. 
The data for each behaviour need to be acquired several times. 
It is laborious for data processing. 
A script is designed to automatically segment the data for this problem. 
Infrared signals are not divisible by m are weed out. 
By entering a sample length m in the script, the data is automatically divided into L$\mid$m samples.

Infrared signals are subject to interference during the acquisition process. 
The interference mainly comes from noise generated by changes in ambient temperature. 
Background subtraction is employed to obtain denoised data. 
Background subtraction is a more widely applied category method in current motion target detection techniques. 
It extracts the target region using the differential operation of different images. 
The difference image is obtained by subtracting the current frame image from a continuously updated background model. 
Motion targets are extracted in the differential images. 
In this paper, the 8 $\times$ 8 IR sensor utilizing AMG8833 consists of an array of 64. 
Human activity is not detected by all 64 arrays at the same time. 
The minimum temperature value in the array can be assumed to be the temperature value of the background model. 
The 64 IR signal values in each frame are sorted from maximum to minimum. 
The array value with the minimum value, $T_{min}$, is selected. 
As mentioned above, $T_{min}$ can be assumed as the ambient temperature of the current frame, $i.e.$, the background model. As shown in Eqs. (1):
\begin{equation}
\label{eq1}
H_a=H_b-T_m
\end{equation}
Where, $H_b=\left(\begin{array}{ccc}\boldsymbol{T}_{11} & \cdots & \boldsymbol{T}_{18} \\ \vdots & \ddots & \vdots \\ \boldsymbol{T}_{81} & \cdots & \boldsymbol{T}_{8 n}\end{array}\right)$,
$T_m=\left(\begin{array}{ccc}\boldsymbol{T}_{\min } & \cdots & \boldsymbol{T}_{\min } \\ \vdots & \ddots & \vdots \\ \boldsymbol{T}_{\min } & \cdots & \boldsymbol{T}_{\min }\end{array}\right)$
are the before processing and background model, respectively.
$H_{a}$ is the IR signal of each frame after background subtraction processing.

\begin{figure*}[!t]
	\centering
	\setcounter{figure}{2}
	\includegraphics[width=6.5in]{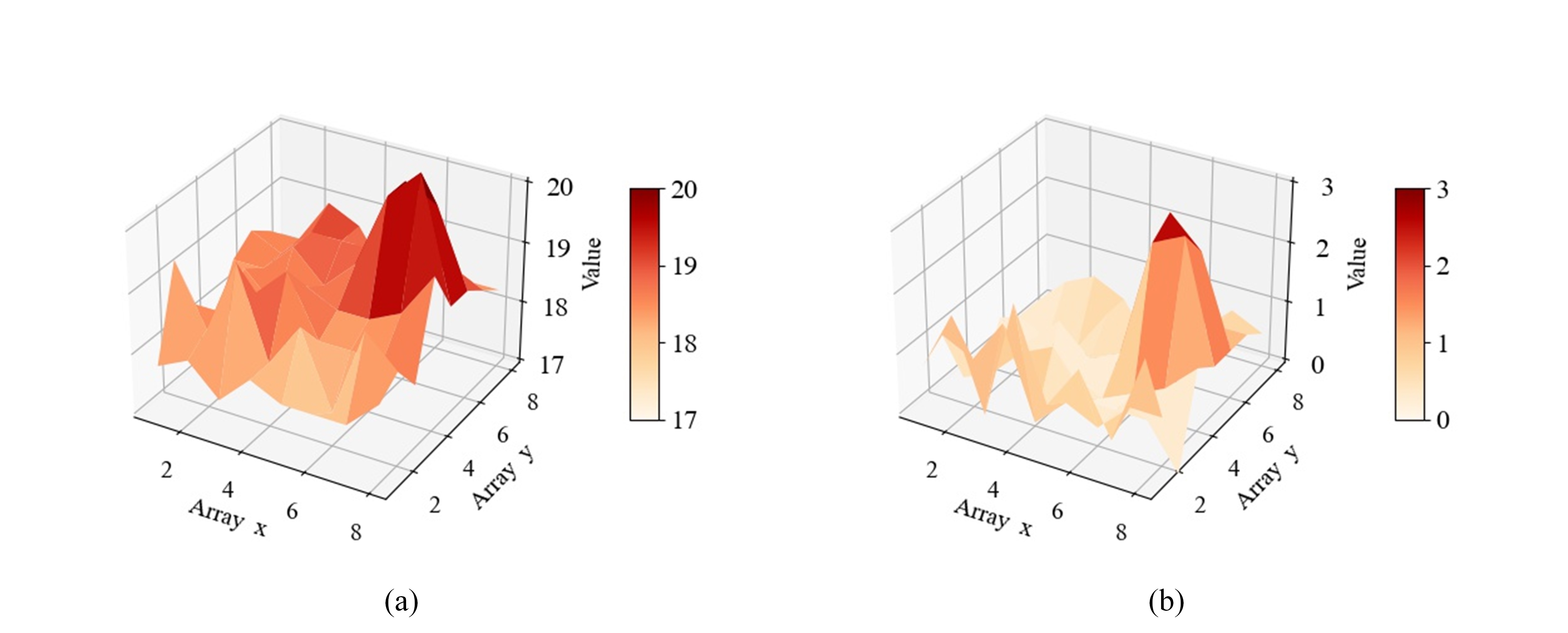}%
	\caption{The heat map comparison of infrared signals before and after data processing. (a) The heat map of the original infrared signal. (b) The heat map of the infrared signal after noise reduction.}
	\label{fig_3}
\end{figure*}
Fig. \ref{fig_3}(a) shows a 3D hotspot map of a frame from raw infrared signal data in standing, and Fig. \ref{fig_3}(b) shows the hotspot map after background subtraction processing on the basis of Fig. \ref{fig_3}(a). 
The comparison shows the infrared signal after data processing has more prominent features.
The noise reflected of other objects, such as fluctuations in the sensor itself and random noise generated in the environment. 
These noises tend to signal mutations. 
The Butterworth filter is employed as a further processing approach after the data is processed by Background subtraction.
It is proved as an effective filter, the Butterworth filter, to filter out the noise reflected by other objects \cite{6542709}. 
More distinctive features of the data are shown on the processed data. 
The data processed by Background subtraction and Butterworth filter is utilized as input for model training. 
Nevertheless, the data processing approaches described above limited effectiveness and are difficult to apply especially when the environment is highly variable. 
It is necessary to design a cross-domain infrared signal feature extraction approach.

\subsection{Semi Supervised Cross-Domain Neural Networks}
The SCDNN proposed in this paper is a cross-domain transfer learning network improved on the basis of DANN. 
DANN is a domain migration method designed on the Generative adversarial network (GAN) \cite{2014Generative} networks. 
Traditional machine learning ensures the training and test sets have similar distributions, otherwise the trained classifier results in poor performance on the test set. 
The purpose of the proposed method is to map source and target domain data with different distributions into the same feature space. 
Mapping is the process to find some kind of transformation criteria. 
The difference between the source and target domain data in the feature space is increasingly smaller.
The trained classifier in the source domain can be directly utilized to classify the data in the target domain. 
Schematic diagram of SCDNN is shown in Fig. \ref{fig_4}. 
It contains of 3 parts, feature extractor, label classifier and domain discriminator. 
The feature extractor is the orange part in the diagram. 
Its role is to map the data into a specific feature space. 
It aims to achieve both the label classifier to discriminate between active categories and the domain discriminator to confuse whether the data comes from the source or target domain. 
The source and target domain samples are employed as input.

In the training process, the input data is initially fed into the feature extractor to extract feature information. 
It is passed into the domain classifier subsequently. 
In the domain classifier, the information is judged to be from the source or target domain and the loss is calculated. 
During the training process, the domain classifier categorizes the input information into the correct domain as much as possible. 
The feature extractor, in contrast, is trained with the opposite objective as the gradient inversion layer incorporated. 
With the adversarial structure, the training purpose is achieved $i.e.$, it is difficult for the domain discriminator to distinguish which domain the data originates from. 
Meanwhile, the information extracted by the feature extractor is fed into the label classifier for categorization, and the loss is calculated. 
The loss of the labeled classifier is reduced and the accuracy of classification is improved by the back-propagation mechanism. 
The accuracy of classification is well-balanced under the supervised and semi-supervised training.
\begin{figure*}[!t]
	\centering
	\setcounter{figure}{3} %决定图片显示的标号，标号-1即为对应的显示图片顺序
	\includegraphics[width=15cm]{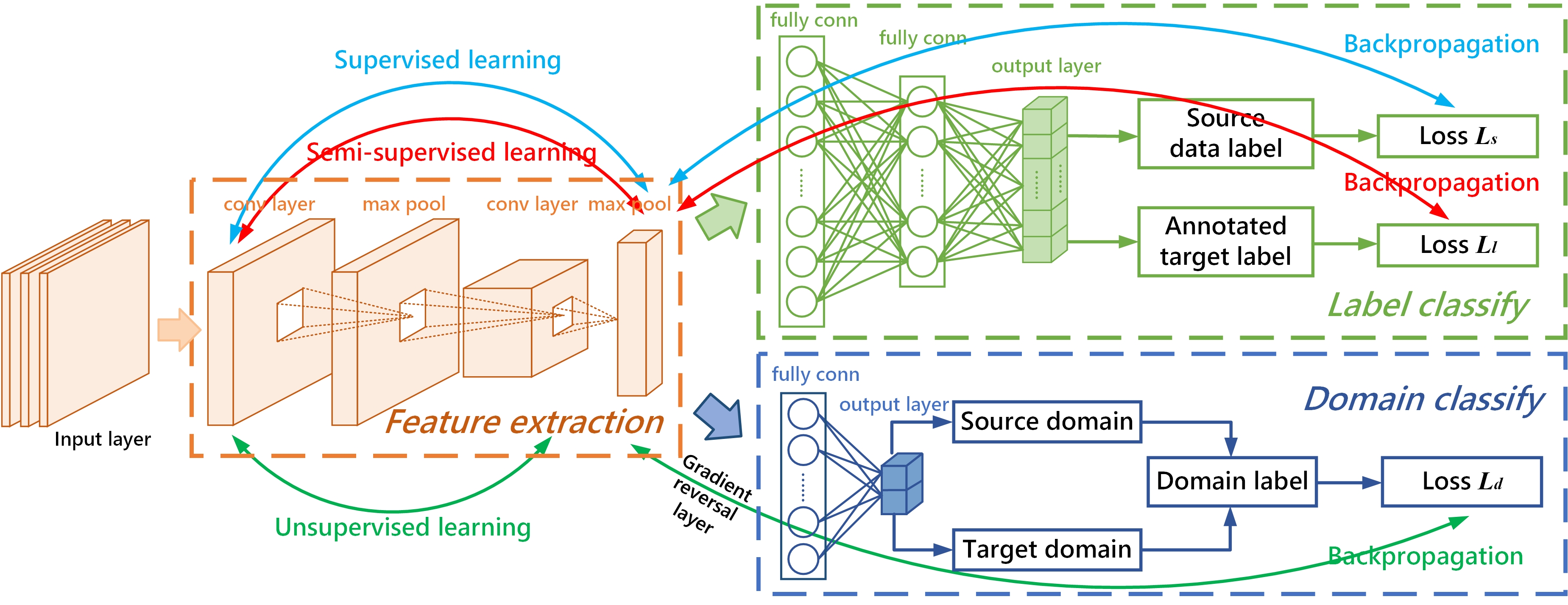}
	\caption{The structure of SCDNN.}
	\label{fig_4}
\end{figure*}
The Sigmoid is utilized as the activation function for the feature extractor and its output is:
\begin{equation}
\label{eq2}
G_{f}({x} ; \mathbf{W}, \mathbf{b})=\operatorname{sigm}(\mathbf{W} {x}+\mathbf{b})
\end{equation}
Where, \emph{x} are the source and target domain samples, \textbf{W} is the weight, and \textbf{b} is the bias term.
The annotated data in the source domain is fed into the feature extractor and the output is fed into the label classifier shown in green. 
The label classifier initially classifies the source domain data and sorts out the correct labels as possible. 
For sample ($x_{s}$, $y_{s}$), $x_{s}$ are the source domain sample, and $y_{s}$ are the label corresponding to $x_{s}$. 
The output of the label classifier can be expressed as:
\begin{equation}
\label{eq3}
G_{l}\left(G_{f}(x_{s}) ; \mathbf{V}, \mathbf{c}\right)=\operatorname{softmax}\left(\mathbf{V} G_{f}(x_{s})+\mathbf{c}\right)
\end{equation}
The activation function of the label classifier is Softmax.
The input $G_{f}(x_{s})$ is the output obtained after the source domain is processed by the feature extractor, \textbf{V} and \textbf{c} are the weights and bias terms respectively. 
The negative log likelihood as loss function, and the loss of the label classifier $L_{l}$ for source domain data is:
\begin{equation}
\label{eq4}
\mathcal{L}_{l}\left(G_{l}\left(G_{f}\left(x_{s}\right)\right), y_{s}\right)=\log \frac{1}{G_{l}\left(G_{f}(x_{s})\right)_{y_{s}}}
\end{equation}
The feature extractor and the label classifier form a feed-forward neural network. 
The optimization objective function for the source domain data is:
\begin{equation}
\label{eq5}
\min _{\mathbf{w}, \mathbf{b}, \mathbf{V}, \mathbf{c}}\left[\frac{1}{n} \sum_{i=1}^{n} \mathcal{L}_{l}^{i}(\mathbf{W}, \mathbf{b}, \mathbf{V}, \mathbf{c})+\lambda \cdot R(\mathbf{W}, \mathbf{b})\right]
\end{equation}
$\mathcal{L}_{l}^{i}(\mathbf{W}, \mathbf{b}, \mathbf{V}, \mathbf{c})$ denotes the label classification loss of the \emph{i}-th sample, $R(\mathbf{W}, \mathbf{b})$ is a regulariser, $\lambda$ is a regularisation parameter. 
$\lambda \cdot R(\mathbf{W}, \mathbf{b})$ aims to prevent the label classification network from overfitting.
The blue part of the diagram shows the domain discriminator. 
Domain classifier is utilized to classify the source of the data (from the source or target domain). 
It is important to note that unsupervised learning is applied here. 
The unlabeled source domain and the unlabeled target domain (no activity category labels, only data source labels) are processed by the feature extractor and fed into the domain discriminator for data source classification. 
The output of the domain discriminator is expressed as:
\begin{equation}
\label{eq6}
G_{d}\left(G_{f}(x_{m}) ; \mathbf{u}, z\right)=\operatorname{sigm}\left(\mathbf{u}^{\top} G_{f}(x_{m})+z\right)
\end{equation}
where $x_{m}$ represents a mixture of samples in the source and target domains, $\mathbf{u}$, $z$ are the weights and bias terms, and the activation function is sigmoid. 
The loss for the domain discriminator $\mathcal{L}_{d}$ is:
\begin{equation}
\label{eq7}
\begin{aligned}
\mathcal{L}_{d}\left(G_{d}\left(G_{f}\left(x_{m}\right)\right), d_{i}\right)&=d_{i} \log \frac{1}{G_{d}\left(G_{f}\left(x_{m}\right)\right)}\\
&+\left(1-d_{i}\right) \log \frac{1}{G_{d}\left(G_{f}\left(x_{m}\right)\right)}
\end{aligned}
\end{equation}
Where, the negative log-likelihood is employed as the loss function.
\begin{figure*}[!t]
	\centering
	\includegraphics[width=6.0in]{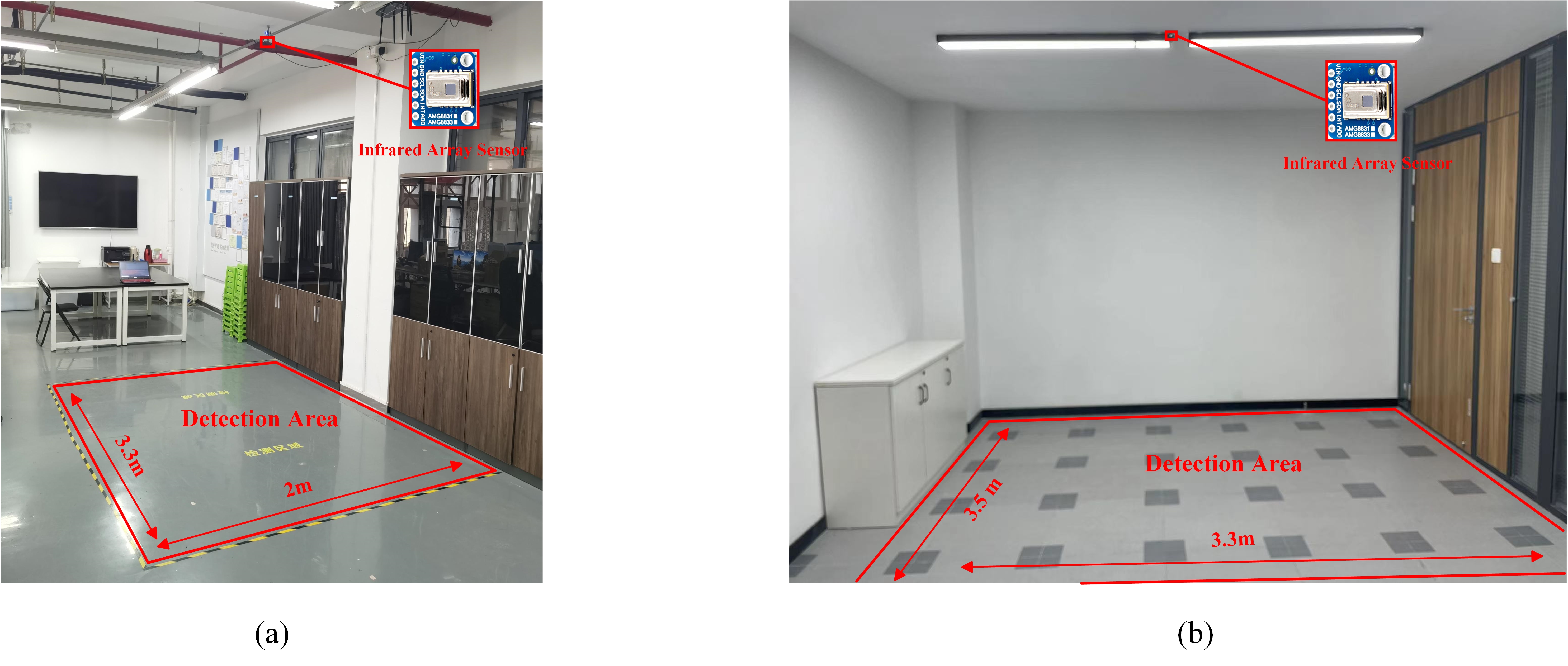}%
	\caption{The layout of the experimental environment. (a) Secne I (Source domain). (b) Secne II (Target domain).}
	\label{fig_5}
\end{figure*}
$d_{i}$ denotes the binary label of the \emph{i}-th sample, which is utilized to indicate whether the sample belongs to the source or target domain. 
The aim is to make it difficult to classify whether the data comes from the source or the target domain. 
In other words to achieve a similar distribution of features between the source and target domain data. 
Feature extractor and domain discriminator are connected by a gradient reversal layer (GRL). 
The GRL automatically reverses the direction of the gradient during back propagation and achieves a identity transform during forward propagation.
If the gradient reversal layer is considered as a function $\mathcal{R}(x)$, its forward and back propagation can be expressed as:
\begin{equation}
\label{eq8}
\mathcal{R}(x)=x
\end{equation}
\begin{equation}
\label{eq9}
\frac{d \mathcal{R}}{d x}=-\mathbf{M}
\end{equation}
where, $\mathbf{M}$ represents an identity matrix. During backpropagation, the gradient of the domain classification loss is automatically inverted before backpropagating to the parameters of the feature extractor, implementing an adversarial loss similar to that of GAN.
The aim of the domain discriminator is to maximise domain classification loss and confuse the target domain data with the source domain data. 
The objective of the label classifier is to minimise activity classification loss and achieve accurate classification on human activity. 
Therefore, the objective function of the domain discriminator is:
\begin{equation}
\label{eq10}
\max _{\mathbf{u}, z}\left[\frac{1}{n} \sum_{i=1}^{n} \mathcal{L}_{d}^{i}(\mathbf{W}, \mathbf{b}, \mathbf{u}, z)\right]
\end{equation}

However, there is a discrepancy between the classes of the source domain and the classes of the target domain. 
In DANN, only unsupervised learning is used to achieve the mapping of the source domain to the target domain data. 
It means that it is difficult for the network to extract features between the different categories of the target domain data. 
The effectiveness of DANN in classifying target domain data is limited.
To solve the above problem, this paper incorporates a semi-supervised learning mechanism in SCDNN. 
A tiny number of target domain samples are labeled for training. 
The annotated target domain samples ($x_{lt}$, $y_{lt}$) are input to the feature extractor for data mapping and the output is obtained as:
\begin{equation}
\label{eq11}
G_{f}\left(x_{lt} ; \mathbf{W}, \mathbf{b}\right)=\operatorname{sigm}(\mathbf{W} x_{lt}+\mathbf{b})
\end{equation}
Then it is fed into the label classifier for classification, giving the output:
\begin{equation}
\label{eq12}
G_{l}\left(G_{f}\left(x_{lt}\right) ; \mathbf{V}, \mathbf{c}\right)=\operatorname{softmax}\left(\mathbf{V} G_{f}\left(x_{s}\right)+\mathbf{c}\right)
\end{equation}
The Loss of label classification for the labeled target domain data is:
\begin{equation}
\label{eq13}
\mathcal{L}_{l}\left(G_{l}\left(G_{f}\left(x_{lt}\right)\right), y_{lt}\right)=\log \frac{1}{G_{l}\left(G_{f}\left(x_{lt}\right)\right)_{y_{lt}}}
\end{equation}
The objective function of $x_{lt}$ is the same as Eqs. (4). The Loss calculated by Eqs. (3) (6) and (12) are both back-propagated through gradient descent to achieve iterative updates of the network parameters.
Labeled classifier in SCDNN benefits from the addition of semi-supervised learning, the features of the target domain samples can be extracted. 
SCDNN enables better distinction the different activities of the target domain.
After the above training steps, test set $x_{tt}$ is fed into the network for testing. 
Test set is mapped in the feature extractor and classified in the label classifier. 
The output after classification is:
\begin{equation}
\label{eq14}
\begin{aligned}
y_{t}&=G_{l}\left(G_{f}\left(x_{tt}\right) ; \mathbf{V}, \mathbf{c}\right)\\
&=\operatorname{softmax}\left(\mathbf{V} G_{f}\left(x_{tt}\right)+\mathbf{c}\right)
\end{aligned}
\end{equation}
The output value $y_{t}$ is a one-dimensional vector with element values corresponding to the probability value of each activity. 
Index of the maximum value in the vector is employed as the final output $i.e.$ the label of the corresponding activity. 
Test set is tested in each training generation. 
The results of the model and tests are saved.
\begin{figure*}[!t]
	\centering
	\includegraphics[width=16cm]{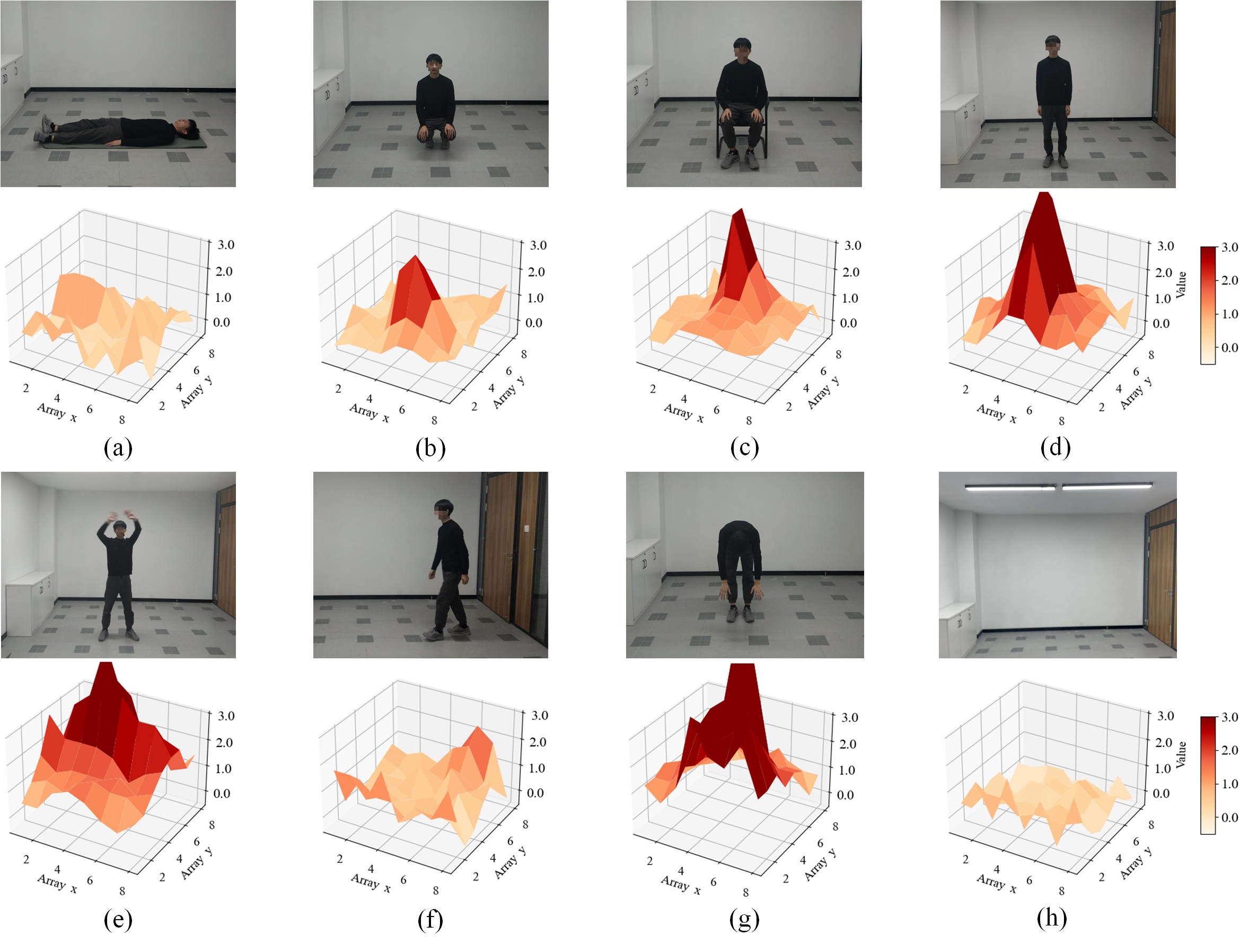}%
	\caption{Five activities in the target domain and the corresponding 3D infrared signal thermograms. \\ (a) Lying. (b) Squatting. (c) Sitting. (d) Standing. (e) Waving. (f) Walking. (g) Stooping. (h) Empty.}
	\label{fig_6}
\end{figure*} 

\section{IMPLEMENTATION AND EVALUATION}
\subsection{Experimental Setup and Data Description}
The method proposed in this paper is validated in two typical indoor environments, as shown in Fig. \ref{fig_5}.
Fig. \ref{fig_5}(a) shows scene I in a relatively empty scene with a detection area of approximately 6.6m$^{2}$, a rectangle of 3.3m and 2m in length and width respectively. 
The low-resolution infrared sensor was mounted approximately 3m above the ground. 
8 types of activities infrared signals were collected as source domain data in scene I. 
Fig. \ref{fig_5}(b) shows scene II , the target domain data was collected here. 
The target domain is an enclosed space of 3.5m in length and 3.3m in width, with an area of approximately 12m$^{2}$. 
The same 8 activities were performed in the target domain as in the source domain and the corresponding infrared signals are acquired. 
The infrared signals were transmitted at 20 frames/s via the I$^{2}$C bus to the ESP8266, and the TPLINK WDR7660 router sends the infrared signals from the ESP8266 to the server for training and testing. 
The hardware configuration of the server as follows: Intel Core i9-7920X CPU, 64G RAM and NVIDIA GeForce GTX 1080Ti GPU. 
Windows 10 operating system was utilized on the server side, a deep learning framework based on Pytorch 1.8.1, and Python 3.6.4 was employed for network construction and programming.

In the experiment, the infrared signals were collected in the source domain (as in Fig. \ref{fig_5}(a)) and the target domain (as in Fig. \ref{fig_5}(b)) for the 8 activities of lying, squatting, sitting, standing, waving, walking, stooping and empty.
Fig. \ref{fig_6} shows the 8 activities in the target domain and the corresponding 3D infrared signal thermograms. 
The infrared signals were acquired by the experimenter at different locations within the detection area. 
The source domain data were collected at 20$^{\circ}$C and the target domain data were collected at 17$^{\circ}$C, 20$^{\circ}$C and 23$^{\circ}$C. 
The experiment was conducted with 4 individuals, 2 males and 2 females.
There were 7,240 acquisitions in the source domain and target domain respectively. 
In the source and target domains, 72,400 frames of IR data were acquired separately. 
Every 10 frames were taken as a sample, with 7,240 samples in the source domain and the same for the target domain.

The parameters of the feature extractor, label classifier and domain discriminator in SCDNN are shown in Tables I, II and III. 
Each contains 10 frames of infrared signal samples converted to $1 \times 20 \times 32$ inputs.
Feature extractor consists of 2 convolutional layers and 2 pooling layers.  
The first convolutional layer has a kernel size of 5 $\times$ 9 and the kernel size of the second convolutional layer is 3 $\times$ 5. 
The activation functions after the convolutional layers are both ReLU. 
Label classifier is connected by 2 fully connected layers and the output layer is 8 neuron nodes corresponding to 8 activities. 
Domain discriminator is connected to the feature extractor by a fully connected layer, and the softmax activation function classifies the source and target domain data. 
The number of training generations is 1000, the batch size is set to 128 and the learning rate is 0.001. To validate the effectiveness of the SCDNN network, the identical target domain test set is fed into the SCDNN and the non-crossed domain model (a normal deep learning model with merely the label classifier module in Table II).

\begin{figure}[t]
	\centering
	\includegraphics[width=8.8cm]{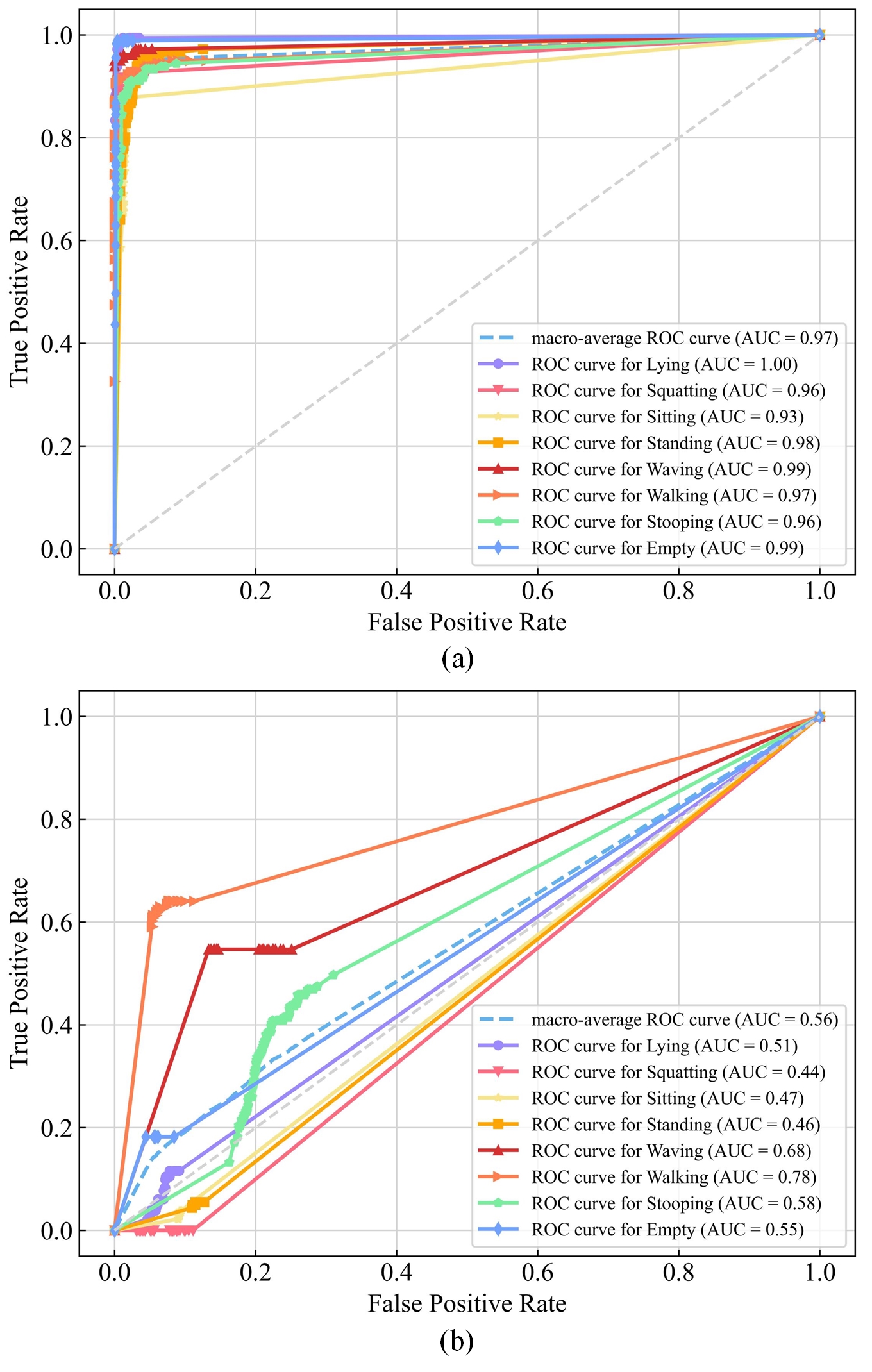}
	\caption{The Receiver-operating-characteristic curve corresponding to the different activities of SCDNN and non-cross-domain approaches in scene II. \\ (a) SCDNN approach. (b) Non-cross domain approach.}
	\label{fig_7}
\end{figure}
\begin{table}[h]
	\begin{center}
		\caption{Parameters for feature extrator.}
		\label{tab1}
		\scalebox{0.8}{		%将表格缩小为原有的0.8倍
			\begin{tabular}{|c|c|c|c|c|}
				\hline
				Layer name                                                       & Input size & Kernel size & Output size & Activation function \\ \hline
				\begin{tabular}[c]{@{}c@{}}Convolutional \\ layer 1\end{tabular} & $1\times20\times32$    & $5\times9$                        & $1280\times16\times24$                       & ReLU                                                          \\ \hline
				\begin{tabular}[c]{@{}c@{}}Max pooling\\ layer\end{tabular}      & $1280\times16\times24$ & $2\times2$                        & $1280\times8\times12$                        & -                                                              \\ \hline
				\begin{tabular}[c]{@{}c@{}}Convolutional \\ layer 2\end{tabular} & $1280\times8\times12$  & $3\times5$                        & $1280\times6\times8$                         & ReLU                                                          \\ \hline
				\begin{tabular}[c]{@{}c@{}}Max pooling\\ layer\end{tabular}      & $500\times6\times8$    & $2\times2$                        & $500\times3\times4$                          &  -                                                             \\ \hline
			\end{tabular}
		}
	\end{center}
\end{table}
\begin{table}[h]
	\begin{center}
		\caption{Parameters for label classifier.}
		\label{tab2}
		\scalebox{0.8}{		%将表格缩小为原有的0.8倍
			\begin{tabular}{|c|c|c|c|}
				\hline
				Layer name                                                         & Input size & Output size                                                & Activation function \\ \hline
				\begin{tabular}[c]{@{}c@{}}Fully connected\\ layer 1\end{tabular}  & 6000       & 1000                                                       & ReLU                \\ \hline
				\begin{tabular}[c]{@{}c@{}}Fully connected \\ layer 2\end{tabular} & 1000       & 500                                                        & ReLU                \\ \hline
				Output layer                                                       & 500        & \begin{tabular}[c]{@{}c@{}}8\\ (8 activities)\end{tabular} & Softmax             \\ \hline
			\end{tabular}
		}
	\end{center}
\end{table}
\begin{table}[!h]
	\begin{center}
		\caption{Parameters for domain discriminator.}
		\label{tab3}
		\scalebox{0.8}{		%将表格缩小为原有的0.8倍
			\begin{tabular}{|c|c|c|c|}
				\hline
				Layer name                                                        & Input size & Output size   & Activation function \\ \hline
				\begin{tabular}[c]{@{}c@{}}Fully connected\\ layer 1\end{tabular} & 6000       & 1000          & ReLU                \\ \hline
				Output layer                                                      & 1000        & \begin{tabular}[c]{@{}c@{}}2\\ (2 domains)\end{tabular} & Softmax             \\ \hline
			\end{tabular}
		}
	\end{center}
\end{table}
\subsection{Evaluation of SCDNN} %Semi-Supervised Cross-Domain \\ Neural Networks
\begin{figure}[t]
	\centering
	\includegraphics[width=8.7cm]{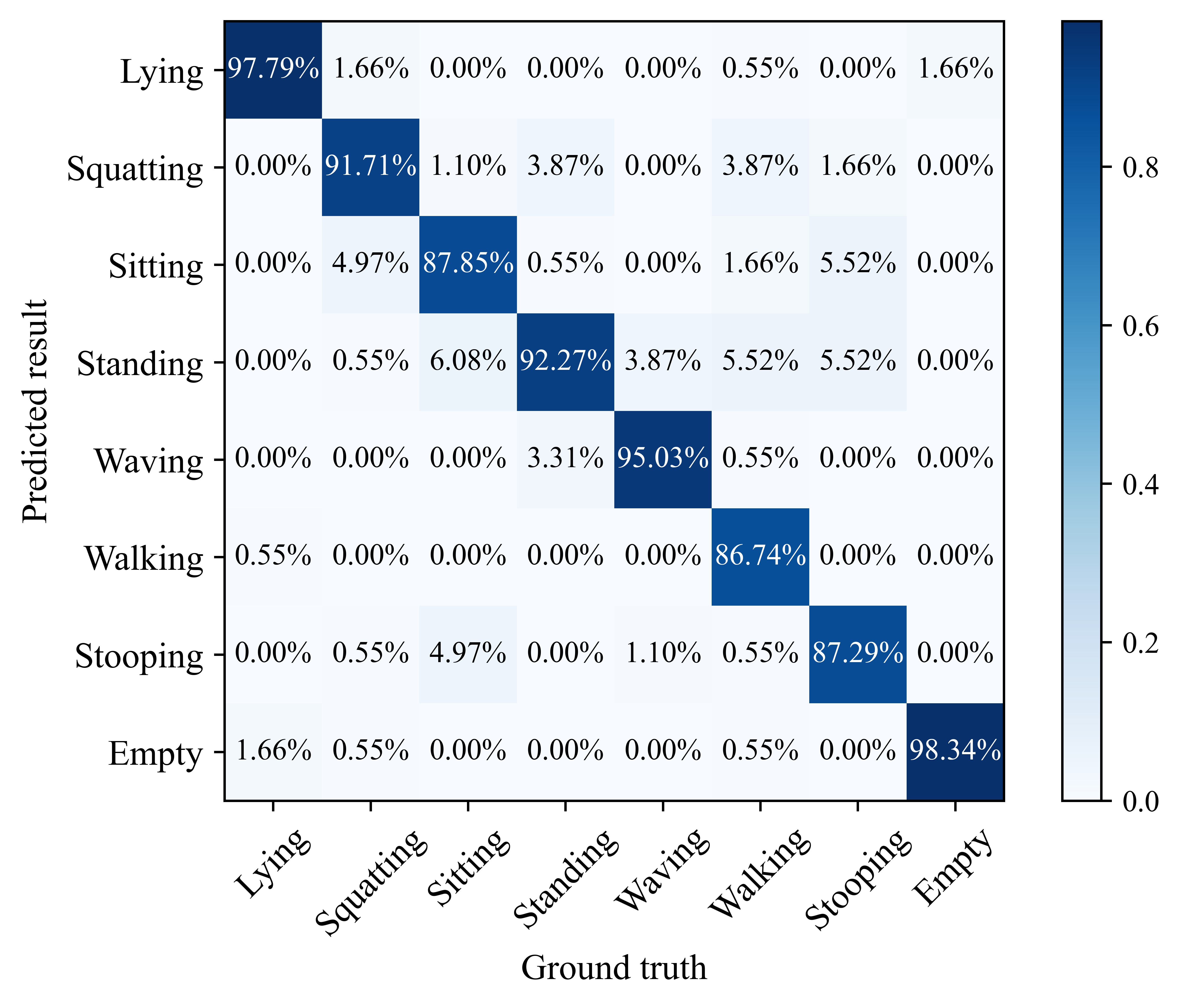}
	\caption{Confusion matrix for the test results of \\ the SCDNN in scene II.}
	\label{fig_8}
\end{figure}

For the target domain samples, 4 samples are randomly selected from the 8 activities to be labeled, 3,480 as unlabeled samples and 1,488 as test samples.
The model trained on the labeled data in scene I is tested utilizing the test set in scene II.
They are trained on 1,000 epochs.
The receiver operating characteristic curve (ROC) is an assessment metric utilized to evaluate the sensitivity and specificity of the model.
The area below the ROC curve is defined as the AUC, and the area closer to 1.0 indicates that the method is more realistic and the model is more effective. 
As shown in Fig. \ref{fig_7}, the ROC of SCDNN is compared with the non-cross domain approach. 
In Fig. \ref{fig_7}(a), the AUC values of all 8 human activities exceed 0.93, and the micro-average reaches 0.97. 
Conversely, the AUC of the non-cross domain approach even fails to exceed 0.6 for all activities except waving and walking. 
It shows that the non-cross domain approach performs extremely weakly in the strange scenario (Scene II). 
The SCDNN still achieves accurate recognition for the 8 activities in cross-domain scenarios with strong robustness. 
Fig. \ref{fig_8} shows the confusion matrix results for the test set. 
As can be seen from the figure, it can be seen that the accuracy for lying, squatting, standing, waving and empty in the test set reaches 97.79\%, 91.71\%, 92.27\%, 95.03\% and 98.34\% respectively. 
The accuracy for sitting, walking and stooping achieved around 87\%.
The average accuracy for the 8 activities reached 92.12\%. 
Most noteworthy, compared to the 5,944 samples utilize in the source domain, only 3,480 unlabeled samples in total and only 4 labeled samples for each activity are taken in the target domain. 
Despite the use of a small-scale target domain data, a high accuracy of HAR is achieved. 
The above results show that SCDNN recognizes the cross-domain activities of human and achieved transfer from the source domain to the target domain. 
High accuracy HAR in target domain is obtained at a lower cost.

\subsection{Impact of unlabeled target domain data}
\begin{figure}[t]
	\centering
	\includegraphics[width=8.3cm]{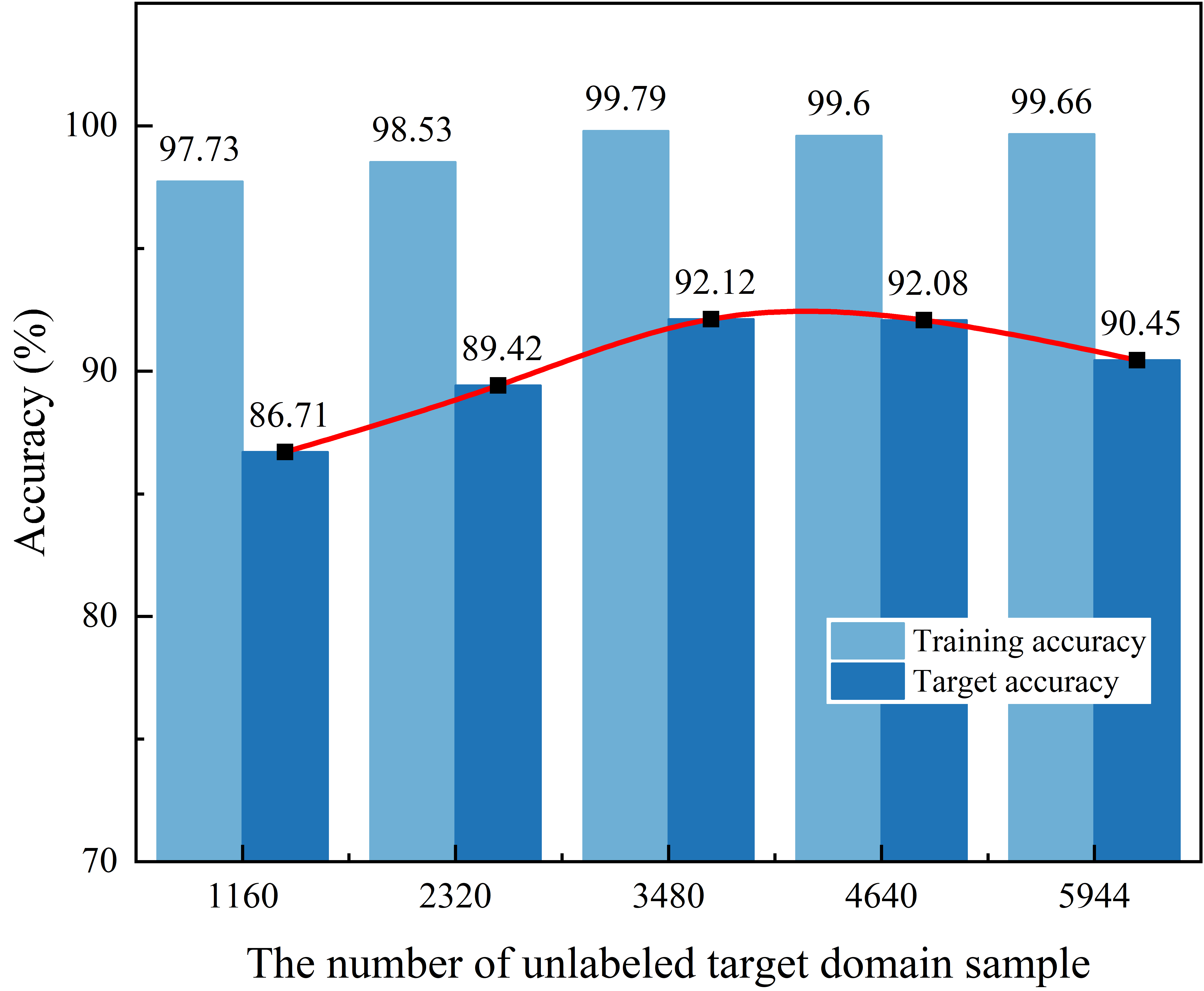}
	\caption{The relationship between number of unlabeled target domain data and recognition accuracy.}
	\label{fig_9}
\end{figure}

The number of unlabeled target domain samples is a key parameter in SCDNN. 
Different numbers of unlabeled samples are selected for experiments to verify the impact of the number of unlabeled samples.
The number of labeled target domain samples is fixed at four for each activity to ensure the uniqueness of the variables. 
As shown in Fig. \ref{fig_9}, the accuracy of the training set is over 97\%. 
As the number of unlabeled samples increased from 1,160 to 3,480, the accuracy of the test set improved from 86.71\% to 92.12\%. 
The accuracy of the test set reaches its highest as the unlabeled samples at 3,480. Overfitting appeared after continuing to increase the number of unlabeled samples. 
The unlabeled samples are increasing while the test set accuracy is decreasing. 
During the increase from 3,480 to 5,944, the accuracy rate decreased from 92.12\% to 90.45\%. The expansion of training samples also leads to an increase in training costs. 
The results implicate a small-scale of unlabeled samples achieve accurate HAR. 
It is appropriated to choose 3,480 unlabeled target domain samples for training.

\subsection{Impact of labeled target domain data}
Similar to the unlabeled samples, the number of labeled samples, important part of SCDNN, is worth exploring. 
Different numbers of labeled samples are employed as experimental subjects to research their effects on experimental results. 
To fix the remaining variables, the number of unlabeled samples is set to 3,480. 
The number of randomly selected labeled target domain samples from each activity are: 0, 2, 4, 6, 8 and 10, and the results are shown in Table IV. 
The precision of the test set improves from 66.79\% to 79.58\% to 92.48\% as the number of labeled samples increase from 0 to 4. 
It can be seen that there is a significant difference in the results between samples with and without labels. 
With the addition of the labeled samples, the effect on the results are clearly positive. 
As described in Chapter III, the model is enabled to obtain the variability between the different activities in the target domain and optimised the network due to the inclusion of the labeled samples. 
It is the unlabeled samples do not have. From Table IV, it can be observed that the Accuracy, Recall and F1-score are higher with the labeled sample compared to the unlabeled sample. 
With 4 labeled samples, the three evaluation indicators are relatively balanced and exceed 92\%. 
\begin{figure}[t]
\centering
\includegraphics[width=8.5cm]{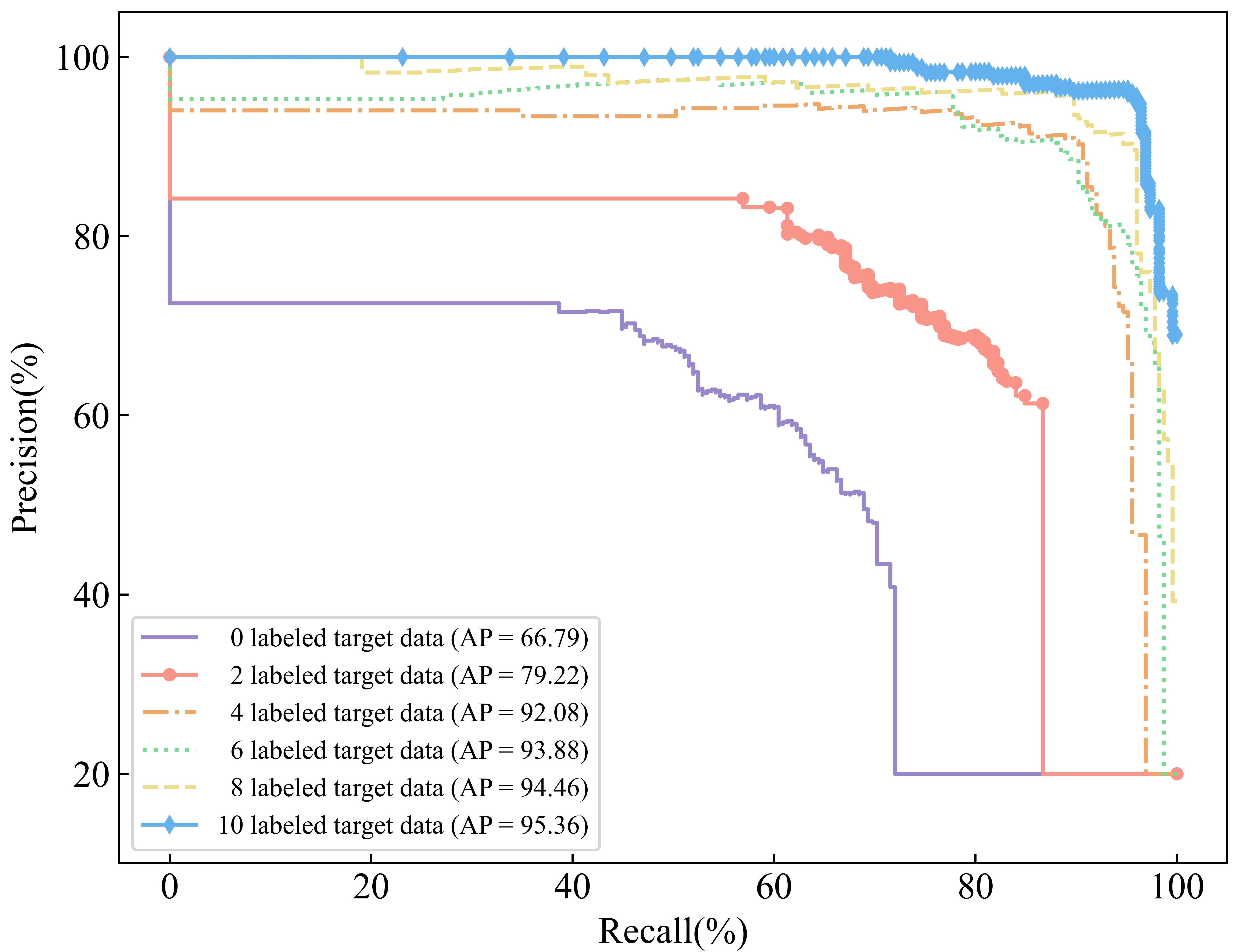}
\caption{The Precision-recall curve in different number of labeled target domain data.}
\label{fig_10}
\end{figure}
These results demonstrate that the performance of the model is excellent for the labeled samples of 4. 
\begin{table}[b]
	\begin{center}
		\caption{Comparison of results with different \\ labeled data.}
		\label{tab4}
		\begin{tabular}{|c|c|c|c|c|c|c|}
			\hline
			\multirow{2}{*}{Performance} & \multicolumn{6}{c|}{Number of labeled data}   \\ \cline{2-7} 
			& 0     & 2     & 4     & 6     & 8     & 10    \\ \hline
			Precision(\%)                & 66.79 & 79.58 & 92.48 & 93.74 & 94.62 & 95.52 \\ \hline
			Recall(\%)                   & 64.23 & 76.87 & 92.13 & 93.65 & 94.54 & 95.27 \\ \hline
			F1-score(\%)                 & 64.50 & 78.90 & 92.17 & 93.63 & 94.16 & 95.38 \\ \hline
		\end{tabular}
	\end{center}
\end{table}
The precision increased from 92.48\% to 95.52\% as the labeled samples are added further up.
The labeled samples and precision show a clear positive correlation. 
Increasing the labeled sample from 4 to 10 limited improvement in precision. 
The precision increases by only 3.04\% over the course of the sample growth, asymptotically reaching the extreme value. 
More data being labeled is undoubtedly more costly. 
On the contrary, although the labeled samples only increase from 2 to 4, the precision increases by 12.9\%. 
Most notably, the precision of the test set reach 92.48\% when the labeled samples for each activity are only 4. 
Fig. \ref{fig_10} shows a plot of the Precision-recall (PR) curve corresponding to different numbers of labeled target domain data. 
Average Precision (AP) is computed as the area under the PR curve. 
AP is utilized to evaluate the comprehensive performance for different numbers of labeled target domain data in the SCDNN framework. 
The APs over 92\% for all cases with the labeled data exceeding 2. 
It indicates that the proposed approach yields a good trade-off between precision and recall with high AP for labeled target domain data more than 2. 
And APs are not significantly distinct from 4 to 10. 
In conclusion, the above results illustrate that a small-scale labeled target data are required to accurately recognize human activities in the SCDNN framework. 
For the sake of precision and training cost, it is more appropriate to 4 labeled samples for each activity.
% \usepackage{multirow}

\subsection{Comparison of methods}
The proposed method in this paper utilizes SCDNN to achieve high accuracy for HAR in target domain. 
To validate the effectiveness of the model, the SCDNN is compared with different algorithms.
K-NearestNeighbor (KNN), CNN, Long short-term memory (LSTM) and DANN are trained and tested on the same data sets.
To demonstrate the advancement of the semi-supervised mechanism proposed in this paper, another semi-supervised learning approach, TSVM, is involved in this paper as a comparison \cite{7041228}. 
The TSVM belongs to the discriminative approach, a typical method in semi-supervised learning. A maximum interval algorithm is used for training both labeled and unlabeled samples to learn the decision boundary. 
The distance interval from the learned classification hyperplane to the nearest sample is maximized. 
Specifically, the TSVM employs a local search strategy to perform iterative solving. 
The unlabeled samples are tagged by the iterated model. 
Thus all samples are labeled. 
The initial model is retrained based on the labeled samples. 
The model is continuously adjusted by finding error-prone samples again.
\begin{figure}[t]
	\centering
	\includegraphics[width=8.3cm]{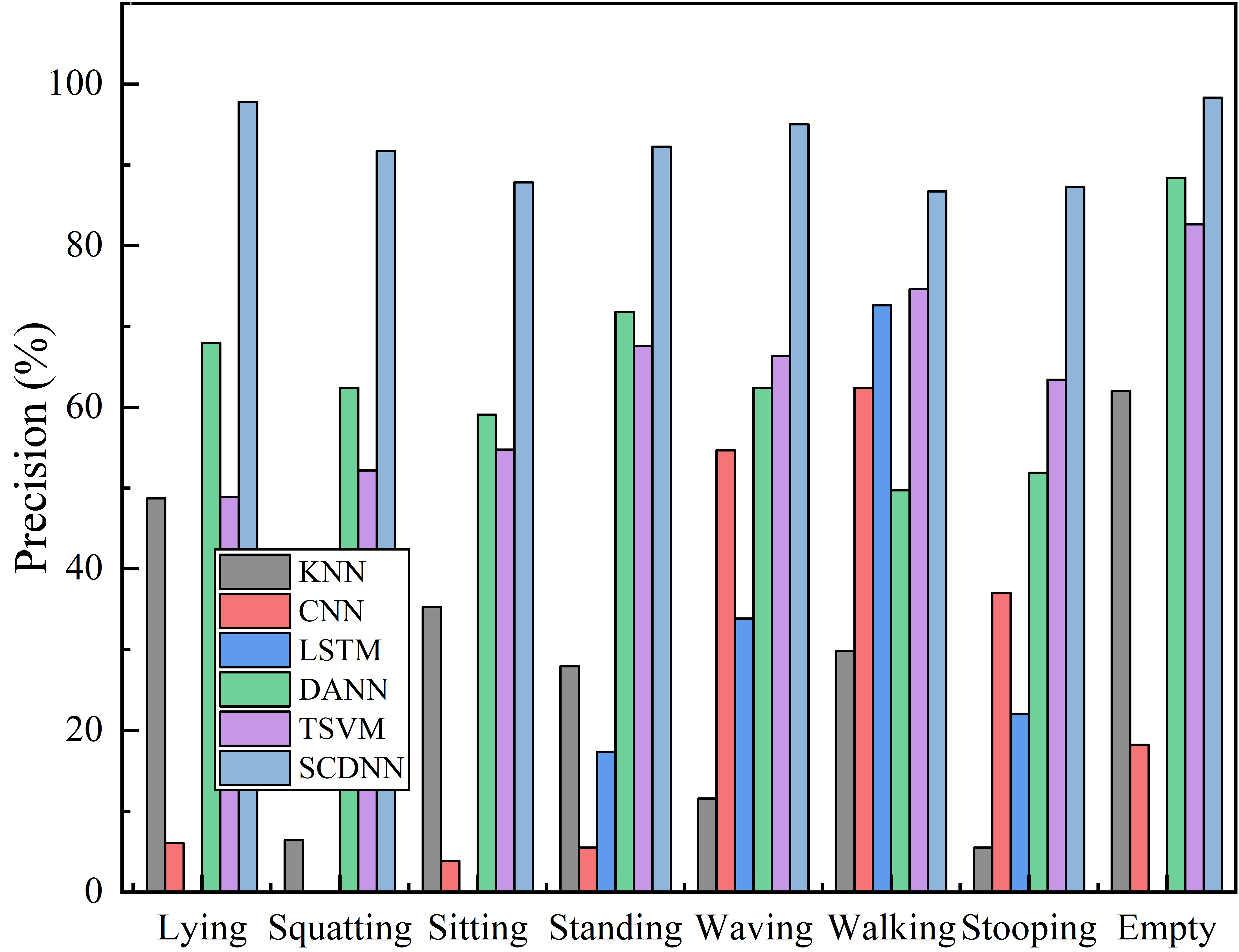}
	\caption{Precision of different algorithms evaluation.}
	\label{fig_11}
\end{figure}

The precision of the different activities for each algorithm was compared in Fig. \ref{fig_11}. 
As shown in the figure, the SCDNN significantly superior precision than the other algorithms in target domain for each activity. 
In Table V, the SCDNN also outperforms the other algorithms in Accuracy, Recall and F1-score. 
Traditional machine learning and conventional deep learning algorithms present poor performance for HAR in target domain. Some activities are even misidentified altogether. 
The overall results are hardly satisfactory. 
Disappointingly, CNN and LSTM performed even worse than traditional machine learning algorithms in crossdomain scenarios. 
It is demonstrated that traditional deep learning algorithms are difficult to be applied in cross-domain scenarios. 
Compared to the traditional methods, the semi-supervised mechanism based TSVM, DANN and SCDNN based on the cross-domain idea perform significantly better than the other algorithms both in overall performance and recognition for most activity.
The TSVM and SCDNN are both semi-supervised learning algorithms. 
Due to the different semi-supervised mechanisms, the semi-supervised of SCDNN approach based on generative adversarial is significantly superior.
Significantly, DANN is an essential improvement over traditional deep learning algorithms and machine learning algorithms. 
From Fig. \ref{fig_11}, although DANN is inferior to SCDNN, the overall recognition is preferable to the other algorithms. 
As environmental changed are complex, unsupervised learning fails to completely eliminate the variability between target and source domain data. 
For unsupervised learning, target domain samples lacking labels are employed as reference. 
It also makes the labeled classifier trained in source domain data not fully applicable to the activity classification in the target domain. 
As a result, the classification effect of DANN on different actions in the target domain is not ideal. 
The SCDNN could effectively extract the features of different activities in target domain by training the labeled target domain data. 
A more accurate HAR is achieved. 
In summary, the SCDNN proposed in this paper is significantly superior to other algorithms both in recognition effect of individual activities and various metrics.
\begin{table}[h]
	\begin{center}
		\caption{Comparison with different methods.}
		\label{tab5}
		\scalebox{0.95}{
			\begin{tabular}{|c|c|c|c|c|c|c|}
				\hline
				Method   & KNN   & CNN   & LSTM  & DANN  & TSVM  & SCDNN \\ \hline
				Accuracy(\%) & 26.32 & 19.51 & 18.86 & 66.23 & 63.82 & 92.12 \\ \hline
				Recall(\%)   & 27.56 & 23.48 & 12.53 & 64.23 & 65.33 & 92.13 \\ \hline
				F1-score(\%) & 26.72 & 20.64 & 13.25 & 64.50 & 62.02 & 92.17 \\ \hline
			\end{tabular}
		}
	\end{center}
\end{table}

\section{CONCLUSION}
In this paper, a novel semi-supervised cross-domain neural network (SCDNN) is proposed for HAR. 
The problem of models trained in the original environment not being applicable to the changed environment is solved. 
Human activity data is captured from an 8 $\times$ 8 low-resolution infrared array sensor.
In the data pre-processing phase, Background subtraction and Butterworth filter are utilized to reduce the noise of the original infrared signal.
The SCDNN is introduced in the training process. 
It achieves domain adaptation by unsupervised learning to align the feature distribution of source and target domain data.
The feature extraction capability of the network is enhanced through training tiny minority labeled target domain data. 
In cross-domain situation, the results indicate that the activities in target domain such as lying, squatting, sitting, standing, waving, walking, stooping and empty can be accurately differentiated with an accuracy of 92.12\%. 
Different amounts of labeled and unlabeled target domain data are experimented with, aiming to achieve the best trade-off between performance and cost. 
Our approach yields a superior performance compared with other deep learning methods. 
The proposed method provides a low-cost and highaccuracy sensing solution for cross-domain application.
There are several limitations of our proposed approach that can be a fruitful direction for further research.
The SCDNN is a cross-domain approach proposed for typical office environments, but its application in other environments is notable.
The suitability of SCDNN needs to be explored at high ambient temperatures.
The compatibility of the system with the outdoor environment is notable to explore.
It might be a good choice to combine the AMG8833 with other sensors for the outdoor environment.
How to utilize a low-resolution infrared sensor for the detection of a group of people is also a very interesting direction to explore. 
The scheme is involved to the future work.

\newpage %另起一页

% your own custom macro containing the $\backslash${\tt{includegraphics}} command to make things
 
\vspace{11pt}

\bf{}\vspace{-33pt}
\begin{IEEEbiography}[{\includegraphics[width=1in,height=1.25in,clip,keepaspectratio]{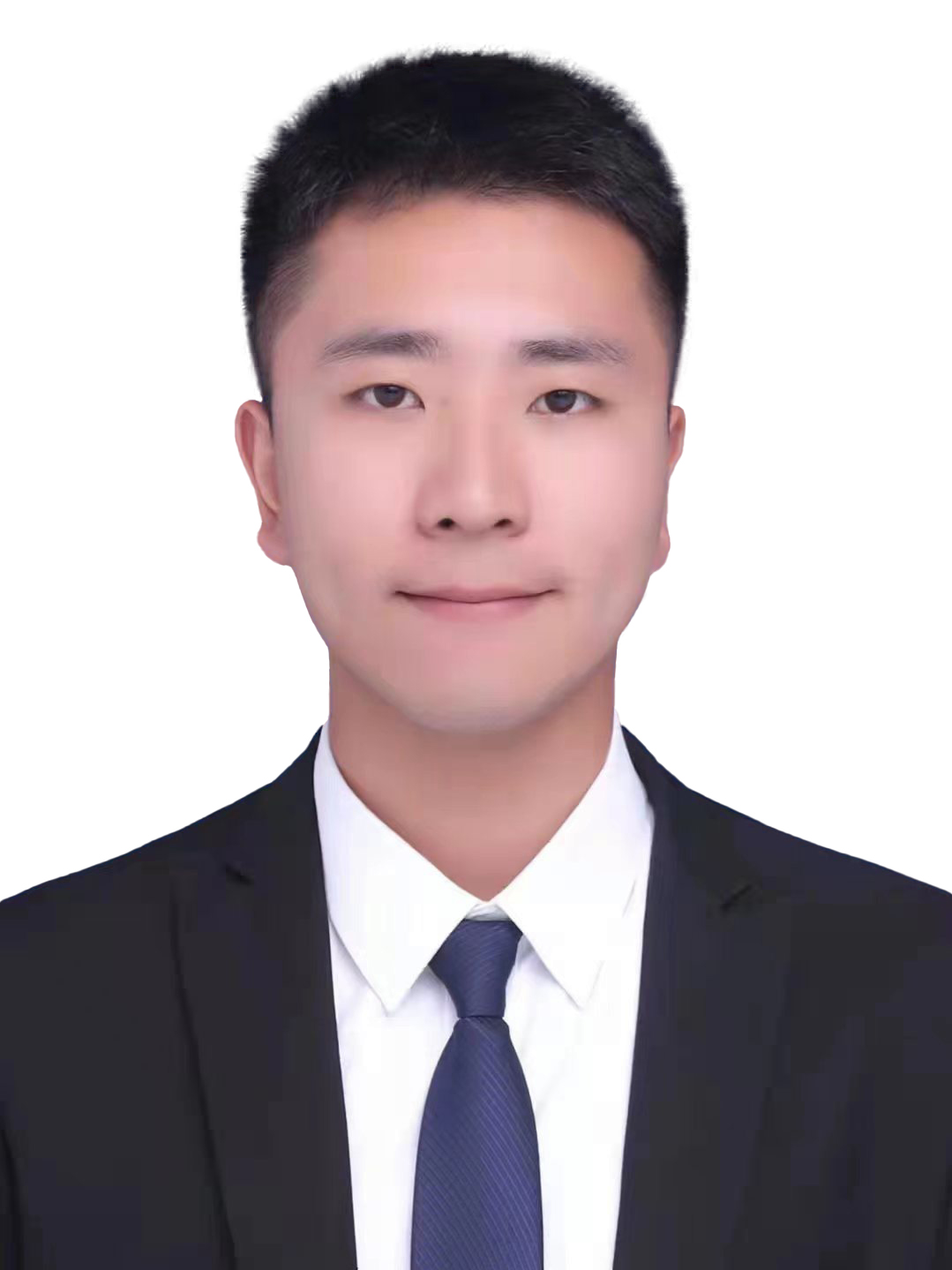}}]{Cunyi Yin}
received the B.S. degree from School of Control Engineering, Chengdu University of Information Technology, Chengdu, China, in 2016, and the M.S. degrees from Fuzhou University, Fuzhou, China, in 2020. He is currently pursuing the Ph.D. degree in  electric machines and electric apparatus with Fuzhou University, Fuzhou, China. His current research interests include human activity recognition, internet of things, wireless sensing and machine learning.
\end{IEEEbiography}

\begin{IEEEbiography}[{\includegraphics[width=1in,height=1.25in,clip,keepaspectratio]{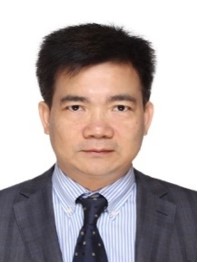}}]{Xiren Miao}
received the B.S. degree from Beihang University, Beijing, China, in 1986, and the M.S. and Ph.D. degrees from Fuzhou University, Fuzhou, China, in 1989 and 2000, respectively, where he is currently a Professor with the College of Electrical Engineering and Automation. His research interests include human activity recognition, electrical and its system intelligent technology, and diagnosis of electrical equipment.
\end{IEEEbiography}

\begin{IEEEbiography}[{\includegraphics[width=1in,height=1.25in,clip,keepaspectratio]{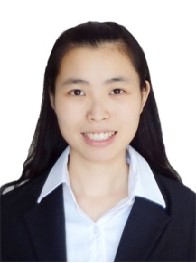}}]{Jing Chen}
received the B.S., M.S., and Ph.D. degrees from Xiamen University, Fujian, China, in 2010, 2013, and 2016, respectively. She is currently an Associate Professor with the College of Electrical Engineering and Automation, Fuzhou University. Her research interests include human activity recognition, intelligent sensor network and machine learning.
\end{IEEEbiography}

\begin{IEEEbiography}[{\includegraphics[width=1in,height=1.25in,clip,keepaspectratio]{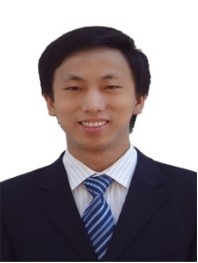}}]{Hao Jiang}
received the B.S. and Ph.D. degrees from Xiamen University, Fujian, China,
in 2008 and 2013, respectively. He is currently an Associate Professor with the College of Electrical Engineering and Automation, Fuzhou University. His research interests include the internet of things, artificial intelligence and machine learning.
\end{IEEEbiography}

\begin{IEEEbiography}[{\includegraphics[width=1in,height=1.25in,clip,keepaspectratio]{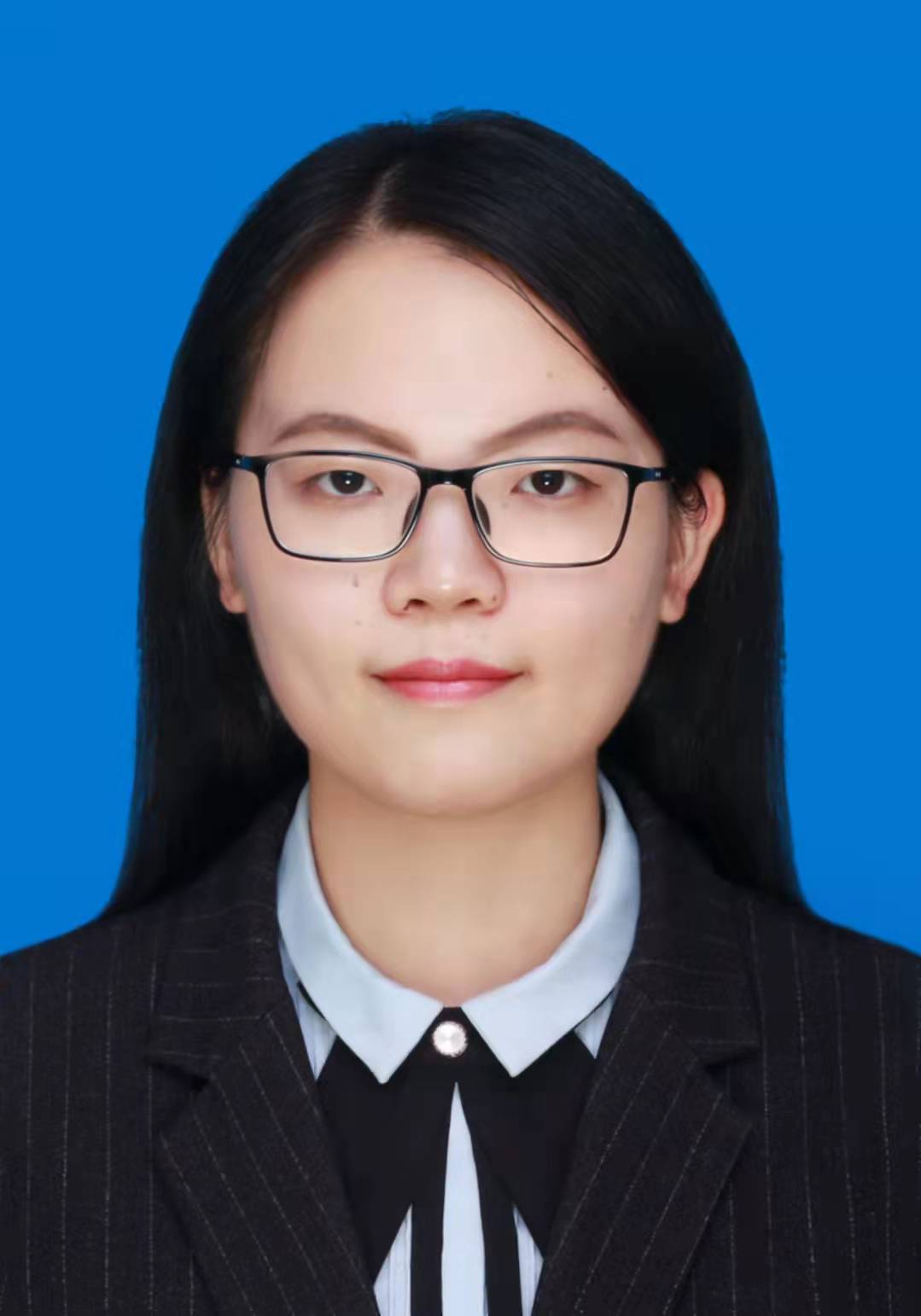}}]{Deying Chen}
received the B.S. degree from College of Electrical Engineering and Automation, Fuzhou University, Fuzhou, China, in 2020. She is currently pursuing the M.E. degree in theory and new technology of electrical engineering with Fuzhou University, Fuzhou, China. Her current research interests include activity recognition and machine learning.
\end{IEEEbiography}

\begin{IEEEbiography}[{\includegraphics[width=1in,height=1.25in,clip,keepaspectratio]{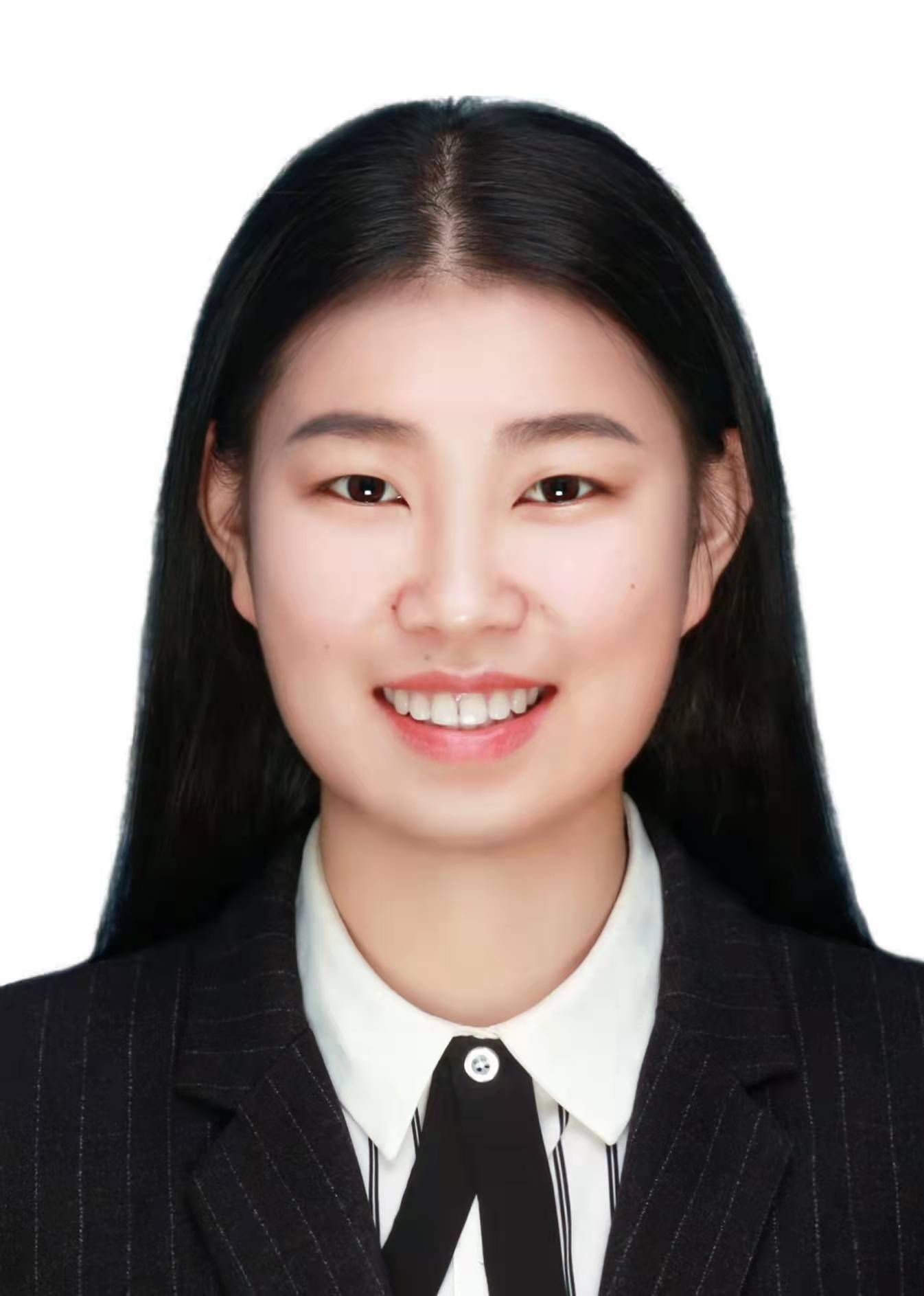}}]{Yixuan Tong}
received the B.S. degree from College of Electrical Engineering and Automation, Fuzhou University, Fuzhou, China, in 2020. She is currently pursuing the M.E. degree in control engineering with Fuzhou University, Fuzhou, China. Her current research interests include activity recognition and machine learning.
\end{IEEEbiography}

\begin{IEEEbiography}[{\includegraphics[width=1in,height=1.25in,clip,keepaspectratio]{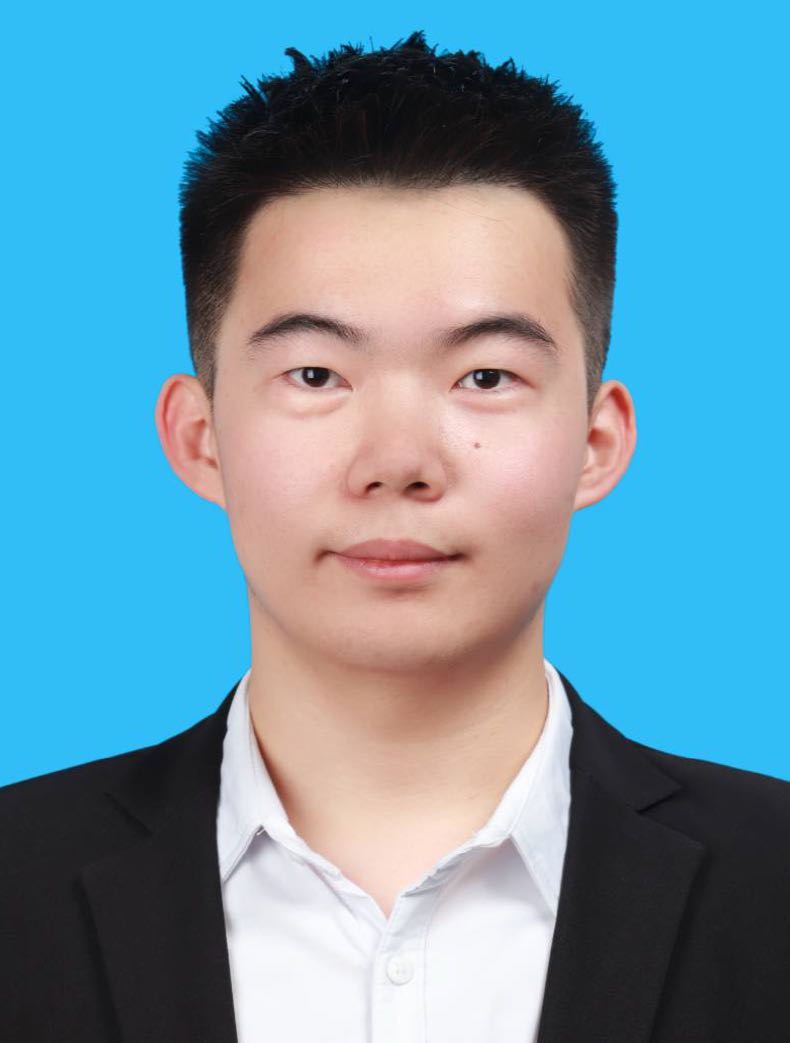}}]{Shaocong Zheng}
received the B.S. degree from College of Electrical Engineering and Automation, Fu zhou University, Fuzhou, China, in 2021. He is currently pursuing the M.E. degree in electronic information with Fuzhou University, Fuzhou, China. His current research interests include the internet of things, sensing networks and artificial intelligence.
\end{IEEEbiography}

\vfill

\end{document}